# Carbon nanotubes as a tip calibration standard for electrostatic scanning probe microscopies

Sergei V. Kalinin,[1] Marcus Freitag,[2] A.T. Johnson,[2] and Dawn A. Bonnell[1*]

[1]*Department of Materials Science and Engineering and Laboratory for Research on the Structure of Matter, University of Pennsylvania, Philadelphia, PA 19104*

[2]*Department of Physics and Astronomy and Laboratory for Research on the Structure of Matter, University of Pennsylvania, Philadelphia, PA 19104*

## ABSTRACT

Scanning Surface Potential Microscopy (SSPM) is one of the most widely used techniques for the characterization of electrical properties at small dimensions. Applicability of SSPM and related electrostatic scanning probe microscopies for imaging of potential distributions in active micro- and nanoelectronic devices requires quantitative knowledge of tip–surface contrast transfer. Here we demonstrate the utility of carbon-nanotube-based circuits to characterize geometric properties of the tip in the electrostatic scanning probe microscopies (SPM). Based on experimental observations, an analytical form for the differential tip-surface capacitance is obtained.

[*] bonnell@lrsm.upenn.edu

In recent years, electrostatic scanning probe microscopies (SPM) such as Electric Force Microscopy (EFM) and Scanning Surface Potential Microscopy (SSPM) have become major tools for the characterization of electric properties of materials on the micron and submicron level.[1] The applicability of these techniques for quantitative nanoscale imaging is hindered by geometric tip effects resulting in smearing of observed potential distributions and cross-talk between potential and topographic images.[2,3,4] For small tip-surface separations tip geometry can be accounted for using the spherical tip approximation and the corresponding geometric parameters can be obtained from electrostatic force- or force gradient distance and bias dependences.[5,6] Such a calibration process is often tedious and tip parameters tend to change with time due to mechanical tip instabilities.[7] Alternatively, the tip contribution to measured surface properties can be quantified directly using an appropriate calibration method.[8] If known, a tip-surface transfer function can be used to deconvolute the tip contribution from experimental data and obtain the *exact* surface potential distribution. Recently, systems with well defined metal-semiconductor interfaces have been considered as a "potential step" standard.[9] However, the presence of surface states and mobile charges significantly affects potential distributions of even grounded surfaces. In addition, such a standard is expected to be sensitive to environmental conditions (humidity, temperature, etc).[10]

Well defined geometry and stability of carbon nanotubes enabled their successful application as SPM probes.[11,12,13,14] Here we propose a carbon nanotube based standard for tip calibration in electrostatic SPM. An ac voltage bias is applied to the nanotube resulting in the oscillation of the SPM tip due to the capacitive force.[15,16]

Taking into account that the typical lateral size of the nanotube is significantly smaller than the tip radius of curvature, the nanotube effectively probes the tip geometry.

Nanotubes are grown by catalytic chemical vapor deposition (CVD)[17,18] directly on a $SiO_2$/Si wafer. Fe/Mo particles on porous alumina act as the catalyst. The nanotubes are grown in an Ar/$H_2$/Ethylene atmosphere at 820°C. This process yields predominantly single wall carbon nanotubes (SWNT) with a small fraction of multiwall nanotubes with a few shells. SWNTs can be distinguished based on the apparent height of 3 nm or less as measured by AFM. The substrate has an oxide layer with a thickness of 225 nm. The degenerately doped silicon acts as a back gate and is grounded. Leads are patterned by e-beam lithography and thermal evaporation of Cr and Au so that the nanotube is a molecular size element in a circuit.

The standard is based on the detection of the amplitude of cantilever oscillation induced by an ac voltage bias ($V_{pp}$ = 200 mV) applied across the carbon nanotube circuit (Fig. 1a). The tip acquires surface topography in the intermittent contact mode and then retraces the surface profile maintaining constant tip-surface separation. Measurements were performed using CoCr coated tips (Metal coated etched silicon probe, Digital Instruments, $l \approx$ 225 µm, resonant frequency ~ 62 kHz) and Pt coated tips (NCSC-12 F, Micromasch, $l \approx$ 250 µm, resonant frequency ~ 41 kHz), further referred to as tip 1 and tip 2. A lock-in amplifier is used to determine the magnitude and phase of cantilever response. The output amplitude, $R$, and phase shift, $\theta$, are recorded by the AFM electronics (Nanoscope-IIIA, Digital Instruments). To avoid cross-talk between the sample modulation signal and topographic imaging, the frequency of ac voltage applied to the nanotube (50 kHz) was selected to be far from the cantilever resonant frequency.

As shown by Jacobs et. al.,[9] the force between the tip and the surface can be written as a function of capacitances as

$$2F_z = C'_{ts}(V_t - V_s)^2 + C'_{ns}(V_n - V_s)^2 + C'_{tn}(V_t - V_n)^2 \qquad (1)$$

where $V_t$ is tip potential, $V_n$ is nanotube potential and $V_s$ is surface potential, $C_{ts}$ is tip-surface capacitance, $C_{ns}$ is nanotube-surface capacitance and $C_{tn}$ is tip-nanotube capacitance. $C'$ refers to derivative of capacitance with respect to the $z$ direction perpendicular to the surface. When an ac bias is applied to the nanotube, $V_n = V_0 + V_{ac}\cos(\omega t)$ and $V_s = V_0$. Therefore, the first harmonic of tip-surface force is:

$$F_{1\omega} = C'_{tn}V_{ac}(V_t - V_0) \qquad (2)$$

In comparison, application of an ac bias to the tip, $V_t = V_{dc} + V_{ac}\cos(\omega t)$ yields

$$F_{1\omega} = C'_{tn}V_{ac}(V_{dc} - V_0) + C'_{ts}V_{ac}(V_{dc} - V_s) \qquad (3)$$

Therefore, applying an ac bias directly to the carbon nanotube allows the tip-surface capacitance to be excluded from the overall force.

Eq. (2) can be generalized in terms of the tip-surface transfer function $C'_z(x,y)$, defined as the capacitance gradient between the tip and a region $dxdy$ on the surface (Fig. 1b)[9] as

$$F_{1\omega} = (V_t - V_0)\int C'_z(x,y)V_{ac}(x,y)dxdy \qquad (4)$$

For the nanotube oriented in the $y$-direction and taking into account small width, $w_0$, of the nanotube compared to the tip radius of curvature, Eq. (4) can be integrated as

$$F_{1\omega}(a) = w_0 V_{ac}(V_t - V_0)\int C'_z(a,y)dy \qquad (5)$$

where $a$ is the distance between the projection of the tip and the nanotube. Assuming a rotationally invariant tip, differential tip-surface capacitance is $C_z(x,y) = C_z(r)$, where $r = \sqrt{x^2 + y^2}$ and Eq.(5) can be rewritten as a function of a single variable, $a$. Therefore, the partial tip-surface capacitance gradient $C'_z(r)$ can be found by numerically solving Eq.(5) using experimentally available force profiles across the nanotube, $F_{1\omega}(a)$.

The validity of the proposed standardization technique is illustrated in Fig. 2. If the measurements are made sufficiently far (1-2 μm) from the biasing contact, the image background and potential distribution along the nanotube are uniform indicating the absence of contact-probe interactions.

Fig. 3 shows topographic and amplitude profiles across the nanotube. The height of the nanotube is 2.7 nm, while apparent width is ~40 nm due to the convolution with the tip shape. Simple geometric considerations yield a tip radius of curvature as $R \approx 75$ nm. Full width at half maximum (FWHM) of the amplitude profile can be as small as ~100 nm and increases with tip-surface separation. This profile is a direct measure of the tip-surface transfer function through Eq. (5).

To analyze the distance dependence and properties of $F_{1\omega}$, amplitude profiles were averaged over ~32 lines and fitted by the Lorentzian function,

$$y = y_0 + \frac{2A}{\pi} \frac{w}{4(x-x_c)^2 + w^2}, \qquad (6)$$

where $y_0$ is an offset, $A$ is area below the peak, $w$ is peak width and $x_c$ is position of the peak. Note that Eq.(6) provides an extremely good description of the experimental data [Fig. 3c]. The offset $y_0$ provides a direct measure of the non-local contribution to the SPM signal due to the cantilever and conical part of the tip.[5,19,20,21] The profile shape is tip dependent and profiles for tip 1 and 2 are compared in Fig. 4d. The distance dependence of peak height $h = 2A/\pi w$ is shown in Fig 4f. For large tip-surface separations $h \sim 1/d$. The distance dependence of width, $w$, is shown in Fig. 4f and is almost linear in distance for $d > 100$ nm. Similar behavior was found for profile width for "potential step" type standards such as ferroelectric domain walls and biased interfaces.[22]

In the particular case of the amplitude profile given by Eq. (6), the local part of the differential tip-surface capacitance can be found solving Eq. (5) as

$$C_z' = \frac{2A}{\pi} \frac{w}{\left(4r^2 + w^2\right)^{3/2}} \tag{7}$$

where $A$ and $w$ are z-dependent parameters determined in Eq.(6) and $r$ is radial distance.

Eq. (7) can be used to determine the tip shape contribution to electrostatic SPM measurements in systems with arbitrary surface potential distributions. For a stepwise surface potential distribution, $V_{surf} = V_1 + (V_2 - V_1)\theta(x)$, where θ(x) is a Heaviside step function, the measured potential profile is $V_{eff} = V_1 + V_2 \arctan(2x/w)/\pi$, provided that the cantilever contribution to the measured potential is small. A similar phenomenological expression is expected to describe phase and amplitude profiles in open-loop SSPM and Scanning Impedance Microscopy (SIM).[23] Fig. 4 shows the phase profile across a grain boundary in a Nb-doped $SrTiO_3$ bicrystal. From independent measurements the double Schottky barrier width is <20 nm, i.e. well below the SPM resolution. Note the excellent agreement between the measured and simulated profile shape. The distance dependence of profile width for the nanotube standard and SIM phase image of grain boundary are compared in Fig. 4, demonstrating excellent agreement. The profile width determined from SSPM measurements is significantly larger indicating feedback and mobile surface charge contribution to the profile width.[24]

To summarize, we have developed a carbon nanotube based standard for the calibration of SPM tips in voltage-modulated SPM. The nanotube standard provides a simultaneous measure of topographic and electrostatic resolution, as well as the convolution function for electrostatic SPM. In contrast to traditional SPM measurements (tip is ac biased) in which the tip interacts both with the dc biased nanotube and the substrate, the latter interaction is effectively excluded. Moreover, surface and oxide trapped charges contribute to the signal for ac tip biasing.[25] Mobile surface charges redistribute under the dc bias, resulting in "smearing" of the potential or electrostatic profile. The characteristic relaxation times for surface charges in air are relatively high and are of order of seconds;[26,27,28] therefore, surface charge dynamics do not contribute to measurements at high (~10-100 kHz) frequencies.

We acknowledge the support from MRSEC grant NSF DMR 00-79909. The authors are grateful to Dr. M. Radosavljevic (UPenn, now at IBM Yorktown Heights) for valuable discussions.

**FIGURE CAPTIONS**

**Fig. 1.** (a) The geometry of the proposed SPM standard. (b) Tip-surface transfer function is defined as capacitance gradient, $C'_z(x,y)$, between the tip and the region *dxdy* located at position *x,y*. Experimentally determined is an integral of $C'_z(x,y)$ [Eq.(5)] as a function of distance from the nanotube, *a*.

**Fig. 2.** Surface topography (a) and Scanning Impedance Microscopy amplitude images (b,c,d) for a carbon nanotube circuit. The contrast is uniform along the tube. Scale is 10 nm (a).

**Fig. 3.** Topographic profile (a) and Scanning Impedance Microscopy amplitude profile (b) across a carbon nanotube. The width of electrostatic profile (~90 nm) is significantly larger than that of the topographic profile (~30 nm), providing a direct measure of tip resolution in topographic and electrostatic measurements. The size of the nanotube per se (~3 nm) is much smaller than either width. (c) Force profiles at lift height of 10 nm (■), 30 nm (▲) and 100 nm (▼) and corresponding Lorentzian fits. (d) Force profiles at lift height of 10 nm, 30nm and 100nm for tip 1 (solid line) and tip 2 (dash line). Peak height (e) and width (f) as a function of tip-surface separation for tip 1 (■) and tip 2 (▲).

**Fig. 4.** Profile width for carbon nanotube standard (▲) and SIM phase image of the SrTiO$_3$ grain boundary (■) as a function of lift height. Inset shows comparison of measured (■) and simulated (line) phase profiles.

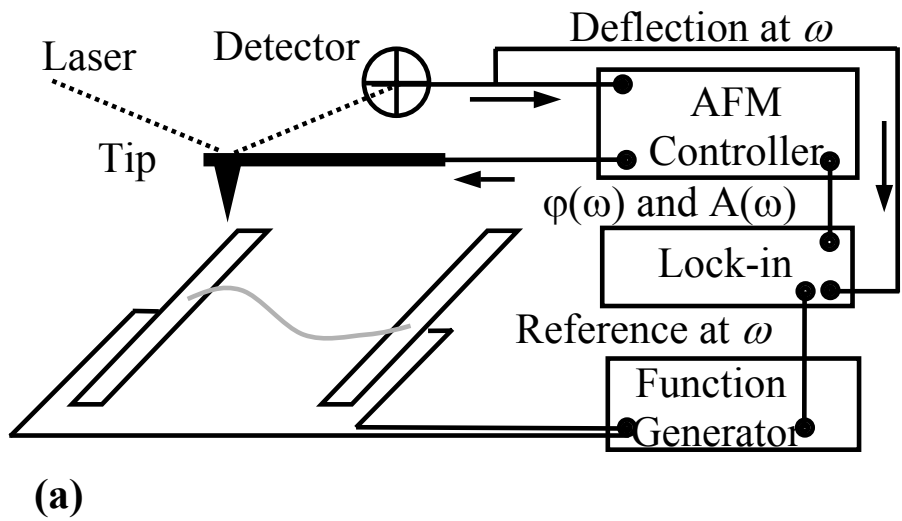

(a)

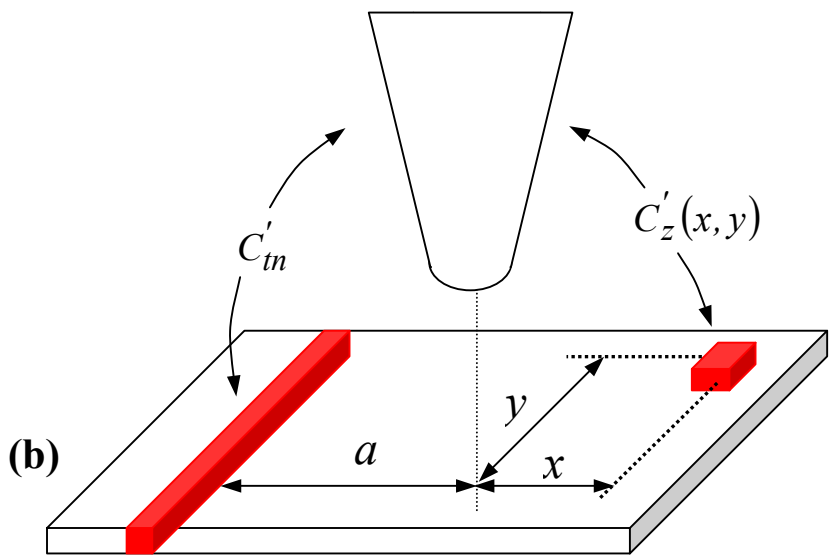

(b)

**Fig. 1.** S.V. Kalinin, M. Freitag, A.T. Johnson, and D.A. Bonnell

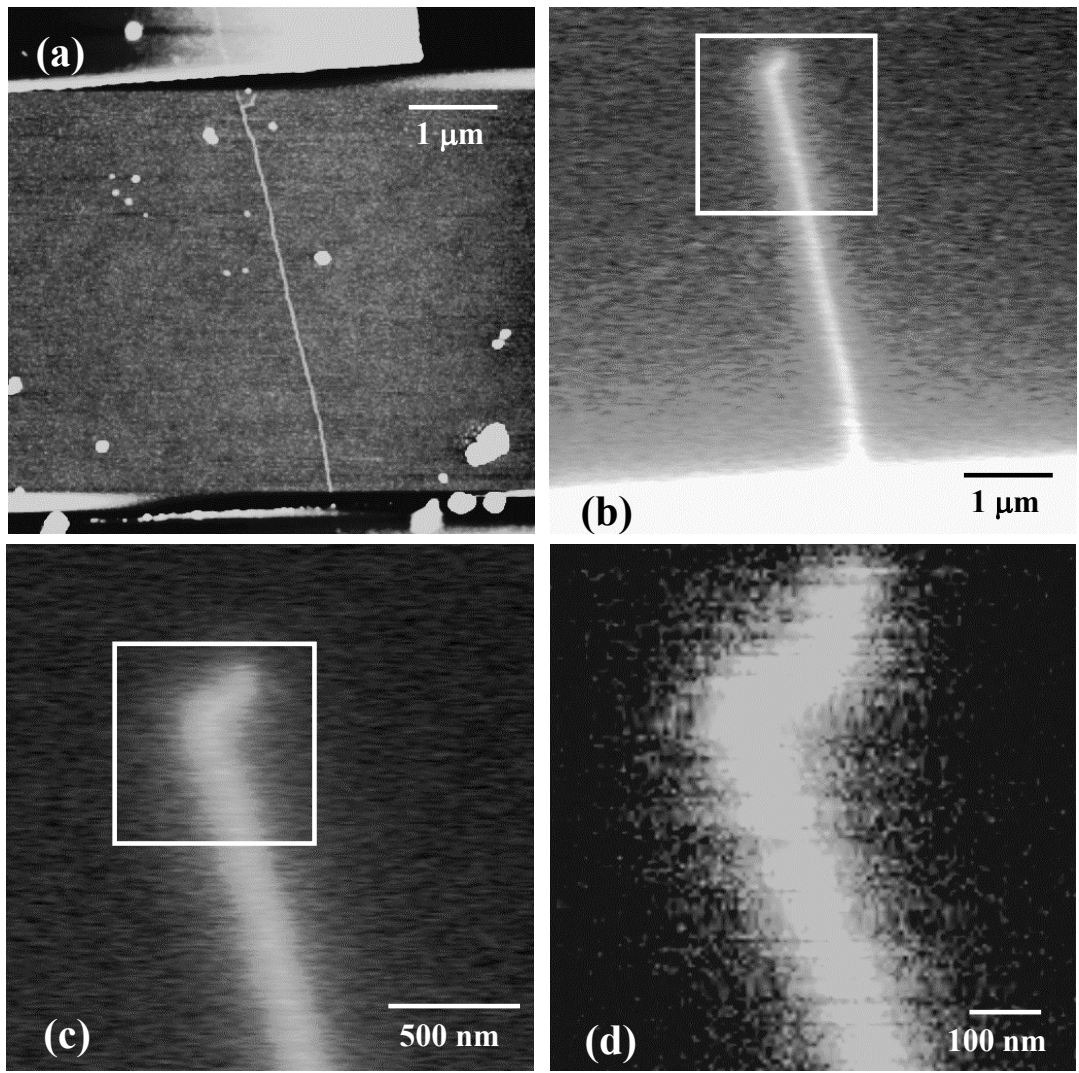

**Fig. 2.** S.V. Kalinin, M. Freitag, A.T. Johnson, and D.A. Bonnell

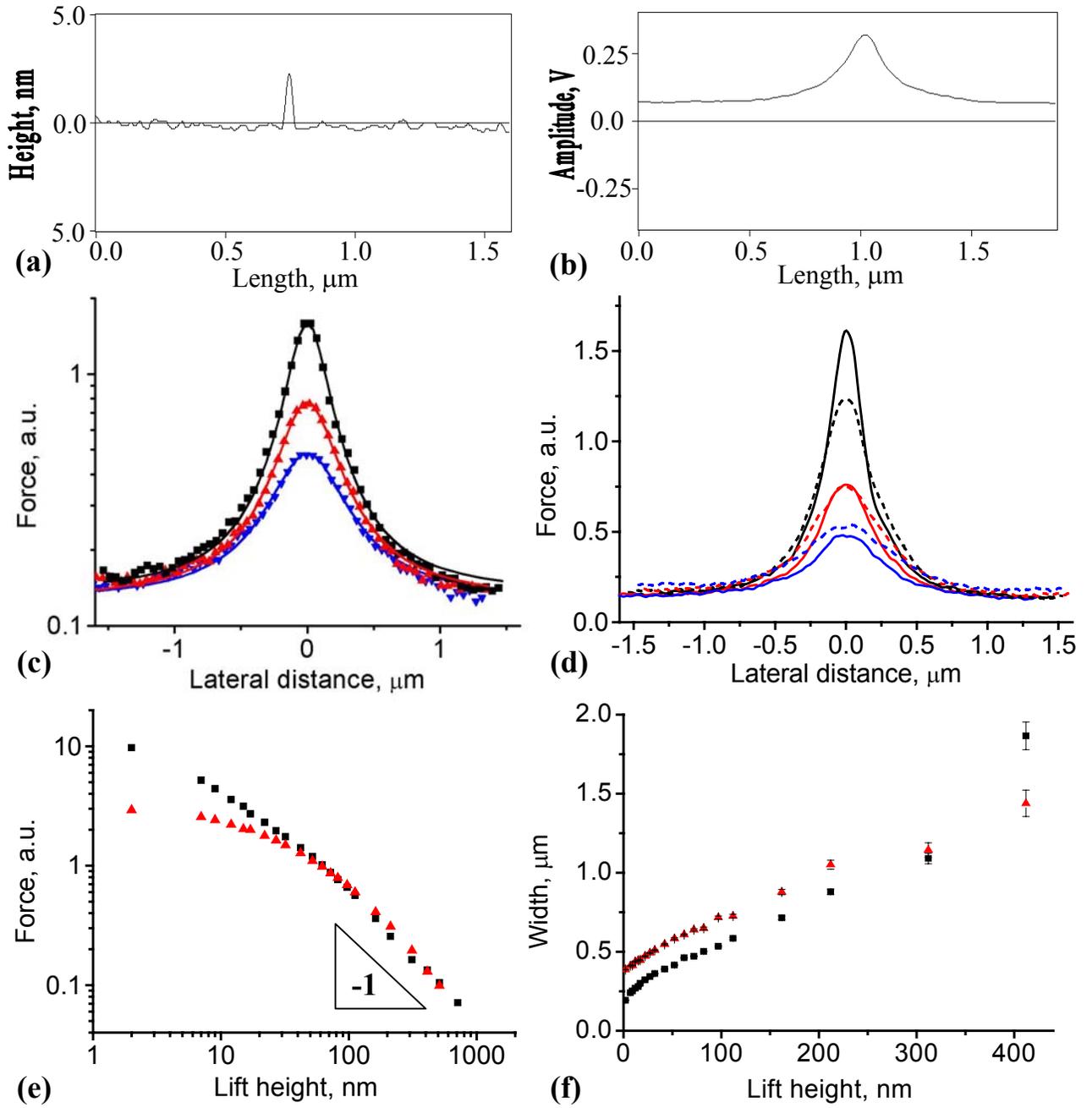

**Fig. 3.** S.V. Kalinin, M. Freitag, A.T. Johnson, and D.A. Bonnell

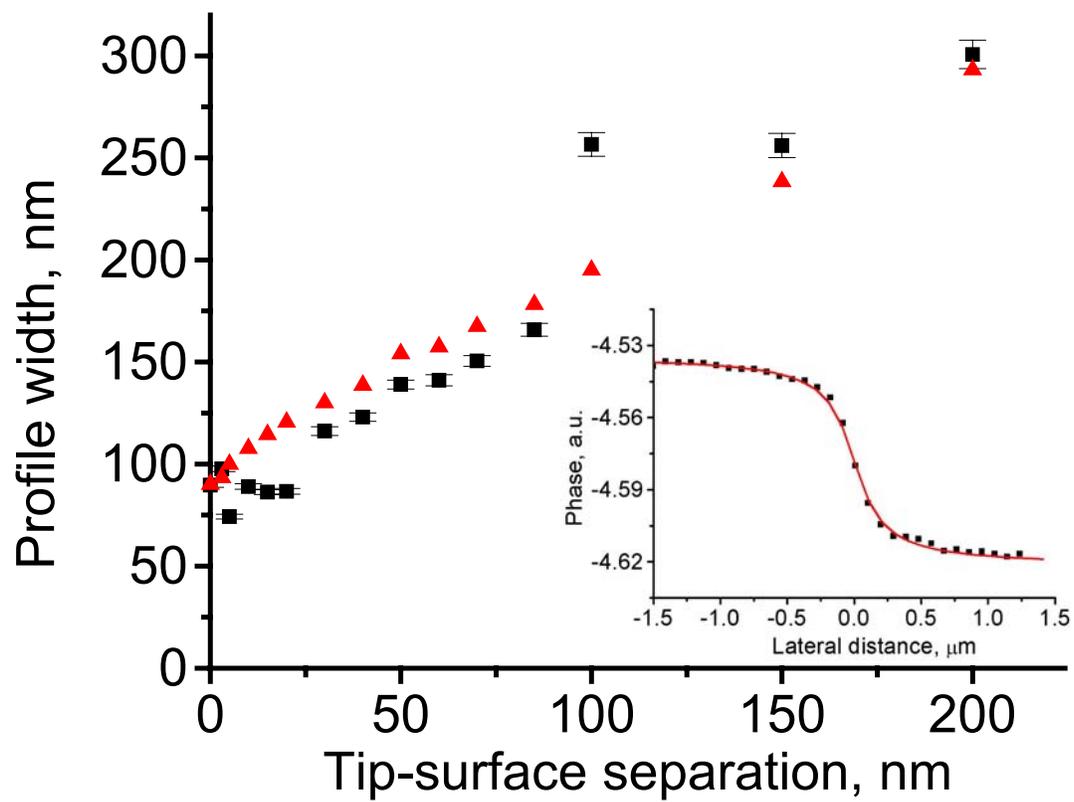

**Fig. 4.** S.V. Kalinin, M. Freitag, A.T. Johnson, and D.A. Bonnell

# REFERENCES

[1] S.V. Kalinin and D.A. Bonnell, in Scanning Probe Microscopy and Spectroscopy: Theory, Techniques and Applications, ed. D.A. Bonnell (Wiley VCH, New York, 2000, p. 205), and references therein

[2] Z.Y. Li, B.Y. Gu, and G.Z. Yang, Phys. Rev. **B 57**, 9225 (1998).

[3] S. Lanyi, J. Torok, and P. Rehurek, J. Vac. Sci. Technol. **B 14**, 892 (1996).

[4] A. Efimov and S.R. Cohen, J. Vac. Sci. Technol. **A 18**, 1051 (2000)

[5] S. Belaidi, P. Girard, and G. leveque, J. Appl. Phys. **81**, 1023 (1997).

[6] L. Olsson, N. Lin, V. Yakimov, and R. Erlandsson, J. Appl. Phys. **84**, 4060 (1998).

[7] H.O. Jacobs, H.F. Knapp, and A. Stemmer, Rev. Sci. Instr. **70**, 1756 (1999)

[8] F. Robin, H. Jacobs, O. Homan, A. Stemmer, and W. Bächtold, Appl. Phys. Lett. **76**, 2907 (2000)

[9] H. O. Jacobs, P. Leuchtmann, O. J. Homan, and A. Stemmer, J. Appl. Phys. **84**, 1168 (1998).

[10] H. Sugimura, Y. Ishida, K. Hayashi, O. Takai, and N. Nakagiri, Appl. Phys. Lett. **80**, 1459 (2002)

[11] H.J. Dai, J.H. Hafner, A.G. Rinzler, D.T. Colbert, R.E. Smalley, Nature **384**, 147 (1996).

[12] S. Takahashi, T. Kishida, S. Akita, and Y. Nakayama Jpn. J. Appl. Phys. **B 40**, 4314 (2001).

[13] S.B. Arnason, A.G. Rinzler, Q. Hudspeth, and A.F. Hebard, Appl. Phys. Lett. **75**, 2842 (1999).

[14] N. Choi, T. Uchihashi, H. Nishijima, T. Ishida, W. Mizutani, S. Akita, Y. Nakayama, M. Ishikawa, and H. Tokumoto, Jpn. J. Appl. Phys. **B 39**, 3707 (2000).

[15] A. Bachtold, M. S. Fuhrer, S. Plyasunov, M. Forero, E.H. Anderson, A. Zettl, and P.L. McEuen, Phys. Rev. Lett. **84**, 6082 (2000).

[16] Sergei V. Kalinin and Dawn A. Bonnell, Appl. Phys. Lett. **78**, 1306 (2001).

[17] J. H. Hafner, C. L. Cheung, and C. M. Lieber, J. Am. Chem. Soc. **121**, 9750 (1999).

[18] M. Freitag, M. Radosavljevic, Y. Zhou, A. T. Johnson, and W. F. Smith, Appl. Phys. Lett. **79**, 3326 (2001)

[19] A.K. Henning, T. Hochwitz, J. Slinkman, J. Never, S. Hoffmann, P. Kaszuba, C. Daghlian, J. Appl. Phys. **77**, 1888 (1995).

[20] Sergei V. Kalinin and Dawn A. Bonnell, Phys. Rev. B **63**, 125411 (2001).

[21] G. Koley, M. G. Spencer, and H. R. Bhangale, Appl. Phys. Lett. **79**, 545 (2001)

[22] Sergei V. Kalinin and Dawn A. Bonnell, unpublished

[23] This result is valid for SPM techniques based on force detection. The resolution in force gradient based techniques such as EFM is better because it is defined by the second derivative of tip-surface capacitance.

[24] Sergei V. Kalinin and Dawn A. Bonnell, Phys. Rev. B **62**, 10419 (2000).

[25] M. Freitag, A.T. Johnson, unpublished.

[26] S. Cunningham, I.A. Larkin, and J.H. Davis, Appl. Phys. Lett. **73**, 123 (1998).

[27] G.H. Buh, H.J. Chung, and Y. Kuk, Appl. Phys. Lett. **79**, 2010 (2001).

[28] T. Tybell, C.H. Ahn, and J.-M. Triscone, Appl. Phys. Lett. **75**, 856 (1999).